\documentstyle[prl,aps]{revtex}

\begin{document}

\title{Comment on ``Halo Properties of the First 1/2$^+$ State in
$^{17}$F from the $^{16}$O(p,$\gamma$)$^{17}$F Reaction''}

\author{K. Riisager}
\address{Institut for Fysik og Astronomi, Aarhus Universitet,
  DK-8000 Aarhus C, Denmark}

\maketitle
\begin{abstract}
 The increase at low energy of the S-factor for the
 $^{16}$O(p,$\gamma$)$^{17}$F$^*$ reaction claimed by Morlock et al.
 to be a new observation is well documented and explained in the literature.
\end{abstract}


Radiative capture to the 5/2$^+$ ground state and 1/2$^+$ first excited
state in $^{17}$F has been studied experimentally 
and theoretically several times, see \cite{rol73,cho75} for a more detailed
account of the early work.
Recently Morlock et al. \cite{mor97} have remeasured the capture reactions,
observed that the S-factor for capture to the first excited state increase
significantly as the energy decreases, performed theoretical calculations and
explained the increase as an effect of the loose binding of the proton
in this state.  (The proton is here in an s$_{1/2}$ orbit around a
$^{16}$O core and is the best documented case for a proton halo state
\cite{rii92,rii94,han95}.)  Morlock et al. stated several times that this
behaviour of the S-factor has not been noticed or explained in any
previous work.  This claim is wrong.

Let us first discuss previous measurements of the S-factors.  The
total S-factor can be derived from measurements of the $^{17}$F
activity produced by a low-energy proton beam, but we shall only
discuss experiments that observe the capture gamma rays and thus give
separate determinations of the capture rates to the two bound states.
In the experiment by Rolfs \cite{rol73} the cross sections were
measured down to beam energies of 300 keV for the excited state
transition and about 600 keV for the ground state transition.  The
former was shown to dominate and to give an S-factor that increases
as the energy decreases (only total S-factors were shown although cross
sections were determined separately for the two transitions, see
figures 7 and 17 in \cite{rol73}).  Calculations were performed
following earlier work \cite{chr61} and reproduced clearly the rise in
the S-factor.  The experiment of Chow et al.\ \cite{cho75} also
separated the two transitions, but did not go to so low beam energies.
Also here calculations were performed that clearly show the rise in
the S-factor at low energies due exclusively to the first excited
state, see their figure 9.  In both papers the direct capture process
is noted to take place mainly into the tail of the bound state wave
function and the enhanced probability for capture into the first
excited state is noted explicitly (figure 4 in \cite{rol73} and page
1684 in \cite{cho75}).  Note that Morlock et al.\ state that the
different energy dependence of the two transitions ``to the best of
our knowledge [\ldots] were not noticed in any previous paper''
although Rolfs and Chow et al.\ both appear in their list of references.  The
recent data \cite{mor97} extend to somewhat lower energies than
\cite{rol73}.  The authors of \cite{mor97} have not done any
comparison with earlier measurements nor with earlier calculations but
a simple reading off the relevant figures indicates that the agreement
is quite good.

The good agreement is also the reason why the adopted rate in
\cite{cau88} for the reaction ``agrees surprisingly well'', in the
words of \cite{mor97}, with the new data.  In contrast to what Morlock
et al.\ state the rate adopted in \cite{cau88} has both experimental
confirmation and has been reproduced theoretically; the expression for
the rate stems from \cite{fow75} where reference is made explicitly to
\cite{rol73}.

The remaining possible new observation in \cite{mor97} is that the
halo properties of the first excited state in $^{17}$F are responsible
for the enhancement of the S-factor at low energies.  However, this
connection (essentially a translation of older insight into new
language) was made \cite{rii92} already five years ago with direct
reference to the proton radiative capture into the $^{17}$F first
excited state.  It was displayed in more detail in a paper on radiative
capture into the $^8$B ground state \cite{rii93}, where figure 1 shows
how the low-energy enhancement is increased drastically as the final
state proton separation energy is decreased.  As explained above, the
physical understanding of this phenomenon was reached fully and stated
clearly already in \cite{rol73,cho75}.  The rather strong statements
to the opposite in \cite{mor97} cannot be substantiated.


\begin{references}
 \bibitem{rol73}C.~Rolfs, Nucl. Phys. \textbf{A217}, 29 (1973)
 \bibitem{cho75}H.C.~Chow, G.M.Griffith and T.H.~Hall, Can. J. Phys.
  \textbf{53}, 1672 (1975)
 \bibitem{mor97}R.~Morlock et al., Phys. Rev. Lett. \textbf{79}, 3837 (1997)
 \bibitem{rii92}K.~Riisager, A.S.~Jensen and P.~M{\o}ller, Nucl. Phys.
  \textbf{A548}, 393 (1992)
 \bibitem{rii94}K.~Riisager, Rev. Mod. Phys. \textbf{66}, 1105 (1994)
 \bibitem{han95}P.G.~Hansen, A.S.~Jensen and B.~Jonson, Ann. Rev. Nucl. Part.
  Sci. \textbf{45}, 591 (1995)
 \bibitem{chr61}R.F.~Christy and I.~Duck, Nucl. Phys. \textbf{24}, 89 (1961)
 \bibitem{cau88}G.R.~Caughlan and W.A.~Fowler, At. Data Nucl. Data Tables
  \textbf{40}, 283 (1988)
 \bibitem{fow75}W.A.~Fowler, G.R.~Caughlan and B.A.~Zimmerman,
  Ann. Rev. Astron. Astrophys. \textbf{13}, 69 (1975)
 \bibitem{rii93}K.~Riisager and A.S.~Jensen, Phys. Lett. \textbf{B301}, 6
  (1993)
\end{references}
\end{document}